\documentclass{article}

\usepackage{amsmath,amssymb,amsfonts} 
\usepackage{color}                    

\usepackage{times}
\usepackage{graphicx} 
\usepackage{subfigure} 

\usepackage{natbib}

\usepackage{algorithm}
\usepackage{algorithmic}
\usepackage{bm}

\usepackage{hyperref}

\usepackage[accepted]{icml2014}

\usepackage{tikz}
\usetikzlibrary{arrows,fit,positioning}
\tikzstyle{block}=[draw opacity=0.7,line width=1.4cm]

\icmltitlerunning{Gaussian Process Volatility Model}

\begin{document} 
\setlength{\belowdisplayskip}{7.5pt} \setlength{\belowdisplayshortskip}{0pt}
\setlength{\abovedisplayskip}{7.5pt} \setlength{\abovedisplayshortskip}{0pt}

\twocolumn[
\icmltitle{Gaussian Process Volatility Model}

\icmlauthor{Yue Wu}{yw289@cam.ac.uk}
\icmlauthor{Jos\'e Miguel Hern\'andez Lobato}{jmh233@cam.ac.uk}
\icmlauthor{Zoubin Ghahramani}{zoubin@eng.cam.ac.uk}
\icmladdress{University of Cambridge, Department of Engineering,
            Cambridge CB2 1PZ, UK}

\icmlkeywords{boring formatting information, machine learning, ICML}

\vskip 0.3in
]

\begin{abstract} 
The accurate prediction of time-changing variances is an important task in the modeling of financial data. 
Standard econometric models are often limited as they assume rigid functional relationships for the variances.
Moreover, function parameters are usually learned using maximum likelihood, which can lead to overfitting.
To address these problems we introduce a novel model for time-changing variances using Gaussian Processes. 
A Gaussian Process (GP) defines a distribution over functions, which allows us to capture highly flexible functional relationships for the variances.
In addition, we develop an online algorithm to perform inference.
The algorithm has two main advantages.
First, it takes a Bayesian approach, thereby avoiding overfitting.
Second, it is much quicker than current offline inference procedures. 
Finally, our new model was evaluated on financial data and showed significant improvement in predictive performance over current standard models.
\end{abstract} 

\section{Introduction} \label{intro}

Time series of financial returns often exhibit heteroscedasticity,
that is the standard deviation or volatility of the returns is time-dependent.
In particular, large returns (either positive or negative) are often followed by
returns that are also large in size. The result is that
financial time series frequently display periods of low and high volatility.
This phenomenon is known as volatility clustering \citep{Cont2001}.
Several univariate models have been proposed for capturing this property.
The best known are the Autoregressive Conditional Heteroscedasticity model (ARCH) \citep{engle1982autoregressive}
and its extension, the Generalised Autoregressive Conditional
Heteroscedasticity model (GARCH) \citep{bollerslev1986generalized}.

GARCH has further inspired a host of variants and extensions
A review of many of these models can be found in \citet{hentschel1995all}.
Most of these GARCH variants attempt to address one or both limitations of GARCH:
a) the assumption of a linear dependency between the current volatility and past volatilities,
and b) the assumption that positive and negative returns have symmetric effects on volatility.
Asymmetric effects are often observed, as large negative returns send measures of volatility soaring, while large positive returns do not \citep{bekaert2000asymmetric,campbell1992no}.

Most solutions proposed in these GARCH variants involve:
a) introducing nonlinear functional relationships for the volatility,
and b) adding asymmetric terms in the functional relationships.
However, the GARCH variants do not fundamentally address the problem that the functional relationship of the volatility is unknown.
In addition, these variants can have a high number of parameters, which may lead to overfitting when learned using maximum likelihood.

More recently, volatility modeling has received attention within the machine learning community, with the development of copula processes \citep{NIPS2010_0784} and heteroscedastic Gaussian processes \citep{Lazaro2011}.
These models leverage the flexibility of Gaussian Processes \citep{rasmussen2006gaussian} to model the unknown relationship in the variances.
However, these models do not address the asymmetric effects of positive and negative returns on volatility.

In this paper we introduce a new non-parametric volatility model, called the Gaussian Process Volatility Model (GP-Vol).
This new model is more flexible, as it is not limited by a fixed functional form.
Instead a prior distribution is placed on possible functions using GPs, and the functional relationship is learned from the data.
Furthermore, GP-Vol explicitly models the asymmetric effects on volatility from positive and negative returns.
Our new volatility model is evaluated in a series of experiments on real financial returns, comparing it against popular econometric models, namely GARCH, EGARCH \citep{nelson1991conditional} and GJR-GARCH \citep{glosten1993relation}.
Overall, we found our proposed model has the best predictive performance.
In addition, the functional relationship learned by our model is highly intuitive and automatically discovers the nonlinear and asymmetric features that previous models attempt to capture.

The second main contribution of the paper is the development of an online algorithm for learning GP-Vol.
GP-Vol is an instance of a Gaussian Process State Space Model (GP-SSM).
Most previous work on GP-SSMs \citep{ko2009gp,deisenroth2009analytic,deisenroth2012expectation} have focused on developing approximation methods for filtering and smoothing the hidden states in GP-SSM, assuming known GP transition dynamics.
Only very recently has \citet{frigola2013bayesian} addressed the problem of learning both the hidden states and the transition dynamics by using Particle Gibbs with ancestor sampling (PGAS) \citep{lindsten2012ancestor}.
In this paper, we introduce a new online algorithm for performing inference on GP-SSMs.
Our algorithm has similar predictive performance as PGAS on financial datasets, but is much quicker as inference is online.


%
\section[Review of GARCH and its variants]{Review of GARCH and GARCH variants} \label{sec:GPReview}
\allowdisplaybreaks

The standard heteroscedastic variance model for financial data is GARCH. 
GARCH assumes a Gaussian observation model \eqref{eq:garchObs} and a linear transition function so that the time-varying variance $\sigma_{t}^{2}$ is linearly dependent on 
$p$ previous variance values and $q$ previous squared time series values:
\begin{align}
	x_{t} &\sim \mathcal{N}(0,\sigma_{t}^{2})\,, \label{eq:garchObs} \\ 
	\sigma_{t}^{2} &= \alpha_{0} + \sum_{j=1}^{q} \alpha_{j} x_{t-j}^{2} + \sum_{i=1}^{p} \beta_{i} \sigma_{t-i}^{2}\,, \label{eq:garch}
\end{align} 
where $x_{t}$ are the values of the return time series being modeled.
The GARCH(p,q) generative model is flexible and can produce a variety of clustering behavior of high and low volatility periods for
different settings of the model coefficients, $\alpha_1,\ldots,\alpha_q$ and $\beta_1,\ldots,\beta_p$.

%
While GARCH is flexible, it has several limitations.
First, a linear relationship between $\sigma_{t-p:t-1}^{2}$ and $\sigma_{t}^{2}$ is assumed.
Second, the effect of positive and negative returns is the same due to the quadratic term $x_{t-j}^{2}$.
However, it is often observed that large negative returns lead to sharp rises in volatility, while positive returns do not \citep{bekaert2000asymmetric,campbell1992no}.

A more flexible and often cited GARCH extension is Exponential GARCH (EGARCH) \citep{nelson1991conditional}:
\begin{align}
	\log(\sigma_{t}^{2}) &= \alpha_{0} + \sum_{j=1}^{q} \alpha_{j} g(x_{t-j}) + \sum_{i=1}^{p} \beta_{i} \log(\sigma_{t-i}^{2})\,, \label{eq:egarch} \\
	g(x_{t}) &= \theta x_{t} + \lambda \abs{x_{t}} \,. \notag
\end{align} 
Asymmetry in the effects of positive and negative returns is introduced through the function $g(x_{t})$.
Then if the coefficient $\theta$ is negative, negative returns will increase volatility.

Another popular GARCH extension with asymmetric effect of returns is GJR-GARCH \citep{glosten1993relation}:
\begin{align}
	\sigma_{t}^{2} = \alpha_{0} +& \sum_{j=1}^{q}\alpha_{j} x_{t-j}^{2} + \sum_{i=1}^{p}\beta_{i} \sigma_{t-i}^{2} + \sum_{k=1}^{r}\gamma_{k} x_{t-k}^{2} I_{t-k} \,, \label{eq:gjr} \\
	I_{t-k} &= \begin{cases} 0\,, \mbox{if } x_{t-k} \geq 0  \\
					   1\,, \mbox{if } x_{t-k} <0
			\end{cases} \,. \notag
\end{align} 
The asymmetric effect is captured by $\gamma_{k} x_{t-k}^{2} I_{t-k}$, which is nonzero if $x_{t-k}<0$. 

\section{Gaussian Process State Space Models} \label{sec:GPrelatedwork}
GARCH, EGARCH and GJR-GARCH can be all represented as General State-Space or Hidden Markov models (HMM) \citep{baum1966statistical, doucet2001sequential},
with the unobserved dynamic variances being the hidden states.
Transition functions for the hidden states are fixed and assumed to be linear in these models.
The linear assumption limits the flexibility of these models.

More generally, a non-parametric approach can be taken where a Gaussian Process prior is placed on the transition function, so that the functional form can be learned from data.
This Gaussian Process state space model (GP-SSM) is a generalization of HMM.
The two class of models differ in two main ways.
First, in HMM the transition function has fixed functional form, while in GP-SSM it is represented by a GP.
Second, in GP-SSM the states do not have Markovian structure once the transition function is marginalized out, 
see Section~\ref{sec:GPinferenceMethods} for details.

However, the flexibility of GP-SSMs comes at a cost.
Specifically, inference in GP-SSMs is complicated.
Most previous work on GP-SSMs \citep{ko2009gp,deisenroth2009analytic,deisenroth2012expectation} have focused on developing approximation methods for filtering and smoothing the hidden states in GP-SSM assuming known GP dynamics.
A few papers considered learning the GP dynamics and the states, but for special cases of GP-SSMs.
For example, \citet{turner2010state} applied EM to obtain maximum likelihood estimates for parametric systems that can be represented by GPs.
Recently, \citet{frigola2013bayesian} learned both the hidden states and the GP dynamics using PGAS \citep{lindsten2012ancestor}.
Unfortunately PGAS is a full MCMC inference method, and can be expensive computationally.
In this paper, we present an online Bayesian inference algorithm for learning the hidden states $v_{1:T}$, the unknown function $f$, and any hyper-parameters $\theta$ of the model.
Our algorithm has similar predictive performance as PGAS, but is much quicker.

\section{Gaussian Process Volatility Model} \label{sec:GPVol}
We introduce now our new nonparametric volatility model, which is an instance of GP-SSM.
We call our new model the Gaussian Process Volatility Model (GP-Vol):
\begin{align}
	x_{t} &\sim \mathcal{N}(0,\sigma_{t}^{2})\,, \label{eq:gpVol0} \\ 
	v_{t} &:= \log(\sigma_{t}^{2}) = f(v_{t-1},x_{t-1}) + \epsilon \,, \label{eq:gpVol1}  \\
	\epsilon &\sim \mathcal{N}(0,\sigma_{n}^{2})\,. \notag
\end{align}
We focus on modeling the log variance instead of the variance as the former has support on the real line.
Equations \eqref{eq:gpVol0} and \eqref{eq:gpVol1} define a GP-SMM. 
Specifically, we place a GP prior on $f$ and letting $z_{t}$ denote $(v_{t},x_{t})$:
\begin{align}
	f &\sim \mathcal{GP}(m,k) \,. \label{eq:gpVol3}
\end{align} 
where $m(z_{t})$ is the mean function and $k(z_{t}, z_{t}')$ is the covariance or kernel function.
In the GP prior, the mean function encodes prior knowledge of the system dynamics.
For example, it can encode the fact that large negative returns lead to increases in volatility.
The covariance function $k(z_{t},z_{t}')$ gives the prior covariance between outputs $f(z_{t})$ and $f(z_{t}')$,
that is $\Cov(f(z_{t}), f(z_{t}'))=k(z_{t},z_{t}')$.
Note that the covariance between the outputs is a function of the pair of inputs.
Intuitively if inputs $z_{t}$ and $z_{t}'$ are close to each other, then the covariances between the corresponding outputs should be large, i.e. the outputs should be highly correlated.

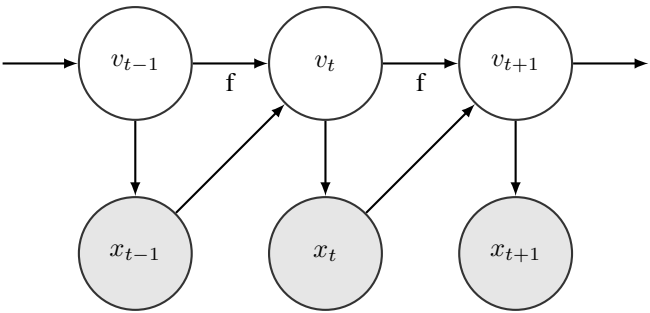
\begin{figure}[h]
\begin{minipage}{.5\linewidth}
\begin{tikzpicture}[scale=1, every node/.style={transform shape}]
\hspace{-1.8cm}
\tikzstyle{first}=[circle, minimum size = 15mm, thick, draw =black!80, node distance = 10mm]
\tikzstyle{main}=[circle, minimum size = 15mm, thick, draw =black!80, node distance = 10mm]
\tikzstyle{main2}=[circle, minimum size = 15mm, thick, draw =black!80, node distance = 10mm]
\tikzstyle{connect}=[-latex, thick]
\tikzstyle{blank}=[circle, minimum size = 15mm, thick, draw =none, node distance = 10mm]
  \node[blank] (x0) {};
  \node[main] (x1) [right=of x0] {$v_{t-1}$};
  \node[main] (x2) [right=of x1] {$v_{t}$};
  \node[main] (x3) [right=of x2] {$v_{t+1}$};
  \node[blank] (xend) [right=of x3] {};
  \node[blank] (z0) [below=of x0] {};
  \node[main2, fill = black!10] (z1) [below=of x1] {$x_{t-1}$};
  \node[main2, fill = black!10] (z2) [below=of x2] {$x_{t}$};
  \node[main2, fill = black!10] (z3) [below=of x3] {$x_{t+1}$};
  \node[blank] (zend) [below=of xend] {};
  \path 
	(x0) edge [connect, thick] (x1) 
	(x1) edge [connect, thick] node [below] {f}  (x2) 
	(x2) edge [connect, thick] node [below] {f} (x3) 
	(x3) edge [connect, thick] (xend)
	(x1) edge [connect] (z1) 
	(x2) edge [connect] (z2) 
	(x3) edge [connect] (z3)
        (z1) edge [connect] (x2) 
	(z2) edge [connect] (x3);
\end{tikzpicture} 
\end{minipage}
\caption{A graphical model for GP-Vol. The transitions of the hidden states $v_{t}$ is represented by the unknown function $f$. $f$ takes as inputs the previous state $v_{t-1}$ and previous observation $x_{t-1}$.}
	\label{fig:gpVol}  
\end{figure}

The graphical model for GP-Vol is given in Figure~\ref{fig:gpVol}.
The explicit dependence of the log variance $v_{t}$ on the previous return $x_{t-1}$ enables us to model asymmetric effects of positive and negative returns on the variance.
Finally, GP-Vol can be extended to depend on $p$ previous log variances and $q$ past returns like in GARCH(p,q).
In this case, the transition would be of the form:
\begin{align}
	v_{t}=f(v_{t-1},v_{t-2},...,v_{t-p},x_{t-1},x_{t-2},...,x_{t-q})+\epsilon \,.
\end{align}

\section{Bayesian Inference for GP-Vol} \label{sec:GPinferenceMethods}
In the standard GP regression setting the inputs and targets are observed, then the function $f$ can be learned using exact inference.
However, this is not the case in GP-Vol, where some inputs and all targets $v_{t}$ are unknown.
Directly learning the posterior of the unknown variables $(f, \theta, v_{1:T})$, where $\theta$ denotes the hyper-parameters of the GP, is a challenging task.
Fortunately, we can target $p(\theta, v_{1:T} | x_{1:T})$, where the function $f$ has been marginalized out.
Marginalizing out $f$ introduces dependencies across time for the hidden states.
First, consider the conditional prior for the hidden states $p(v_{1:T}|\theta)$ with known $\theta$ and $f$ marginalized out. 
It is not Gaussian but a product of Gaussians:
\begin{align} 
	p(v_{1:T} | \theta) &= p(v_{1} |\theta) \prod_{t=2}^{T} p(v_{t} | \theta, v_{1:t-1}, x_{1:t-1}) \label{eq:GPVolPrior} \,.
\end{align}
Each term in Equation~\eqref{eq:GPVolPrior} can be viewed as a standard one-step GP prediction under the prior.
Next, consider the posterior for the states $p(v_{1:t} | \theta, x_{1:t})$.
For clarification the prior distribution of $v_{t}$ depends on all the previous states $v_{1:t-1}$ and previous observations $x_{1:t-1}$.
In contrast the posterior for state $v_{t}$ depends on all available observations $x_{1:t}$.

The posterior $p(v_{t} | \theta, v_{1:t-1}, x_{1:t})$ can be approximated with particles.
We now describe a standard sequential Monte Carlo (SMC) particle filter to learn this posterior.
Let $v_{1:t-1}^{i}$ with $i=1,...,N$ be particles that represent chains of states up to $t-1$ with corresponding weights $W_{t-1}^{i}$.
Then the posterior distribution of $p(v_{1:t-1} | \theta, x_{1:t-1})$ is approximated by weighted particles:
\begin{align}
	\hat{p}(v_{1:t-1} | \theta, x_{1:t-1}) &= \sum_{i=1}^{N} W_{t-1}^{i} \delta_{v_{1:t-1}^{i}}(v_{1:t-1}) \label{eq:particleApprox} \,.
\end{align}
In addition, the posterior for $v_{t}$ can be approximated by propagating the previous states forward and importance weighting according to the observation model.
Specifically, sample a set of parent indices $J$ according to $W_{t-1}^{i}$.
Then propagate forward particles $\{ v_{1:t-1}^{j} \}_{j \in J}$.
Proposals for $v_{t}$ are drawn from its conditional prior:
\begin{align}
	v_{t}^{j} &\sim p(v_{t} | \theta, v_{1:t-1}^{j}, x_{1:t-1}) \label{eq:gpVolProp} \,.
\end{align}
The proposed particles are importance-weighted according to the observation model:
\begin{align}
	w_{t}^{j} &= p(x_{t} | \theta, v_{t}^{j}) \label{eq:gpVolISWeight} \,, \\
	W_{t}^{j} &= \frac{w_{t}^{j}}{\sum_{k=1}^{N} w_{t}^{k}} \label{eq:gpVolNormWeight} \,.
\end{align}
Finally the posterior for $v_{t}$ is approximated by:
\begin{align}
	\hat{p}(v_{t} | \theta, v_{1:t-1}, x_{1:t}) &= \sum_{j=1}^{N} W_{t}^{j} \delta_{v_{t}^{j}}(v_{t}) \label{eq:particleApprox2} \,.
\end{align}

The above setup learns the states $v_{t}$, assuming that $\theta$ is given.
Now consider the desired joint posterior $p(\theta, v_{1:T} | x_{1:T})$.
To learn the posterior, first a prior $p(\theta,v_{1:T})=p(v_{1:T}|\theta)p(\theta)$ is defined.
This suggests that the hyper-parameters $\theta$ can also be represented by particles and filtered together with the states.
Naively filtering $\theta$ particles without regeneration will fail due to particle impoverishment,
where a few or even one particle for $\theta$ receives all the weight.
To resolve particle impoverishment, algorithms, such as the Regularized Auxiliary Particle Filter (RAPF) \citep{liu1999combined}, regenerates parameter particles at each time step by sampling from a kernel.
This kernel introduces artificial dynamics and estimation bias, but works well in practice \citep{wu2013dynamic}.

RAPF was designed for Hidden Markov Models, but GP-Vol is marginally non-Markovian.
Therefore we design a new version of RAPF for non-Markovian systems and refer to it as the Regularized Auxiliary Particle Chain Filter (RAPCF), Algorithm~\ref{alg:rapfGPVol}.
There are four main parts to RAPCF.
First, there is the Auxiliary Particle Filter (APF) part of RAPCF in lines~\ref{lst:line:apf1} and ~\ref{lst:line:apf2}.
The APF \citep{pitt1999filtering} proposes from more optimal importance densities, by considering how well previous particles would represent the current state \eqref{eq:pointEstimateImportanceWeight}.
Second, the more likely particle chains are propagated forward in line~\ref{lst:line:apf3}.
The main difference between RAPF and RAPCF is in what particles are propagated forward.
In RAPCF for GP-Vol, particles representing chains of states $v_{1:t-1}^{i}$ that are more likely to describe the new observation $x_{t}$ are propagated forward, as the model is non-Markovian,
while in standard RAPF only particles for the previous state $v_{t-1}^{i}$ are propagated forward.
Third, to avoid particle impoverishment in $\theta$, new particles are generated by applying a Gaussian kernel, in line~\ref{lst:line:shrinkageKernel}.
Finally, the importance weights are computed for the new states adjusting for the probability of its chain of origin \eqref{eq:realImportanceWeights}.

RAPCF is a quick online algorithm that filters for unknown states and hyper-parameters.
However, it has some limitations just as standard RAPF.
First, it introduces bias in the estimates for the hyper-parameters as sampling from the kernel in line~\ref{lst:line:shrinkageKernel} adds artificial dynamics.
Second, it only filters forward and does not smooth backward.
This means that the chains of states are never updated given new information in later observations.
Consequently, there will be impoverishment in distant ancestors $v_{t-L}$, since these ancestor states are not regenerated.
When impoverishment in the ancestors occur, GP-Vol will consider the collapsed ancestor states as inputs with little uncertainty.
Therefore the variance of the predictions near these inputs will be underestimated.

Many of the potential issues faced by RAPCF can be addressed by adopting a full MCMC approach.
In particular, Particle Markov Chain Monte Carlo (PMCMC) procedures \citep{andrieu2010particle} established a framework for learning the hidden states and the parameters for general state space models.
Additionally, \citet{lindsten2012ancestor} developed a PMCMC algorithm called Particle Gibbs with ancestor sampling (PGAS) for learning non-Markovian state space models,
which was applied by \citet{frigola2013bayesian} to learn GP-SSMs.

\begin{algorithm}[H]
   \caption{RAPCF for GP-Vol}
   \label{alg:rapfGPVol}
\belowdisplayskip=-5pt 
\begin{algorithmic}[1]
   \STATE {\bfseries Input:} data $x_{1:T}$, number of particles $N$, shrinkage parameter $0<a<1$,  priors $p(\theta)$.
   \STATE At $t=0$, sample $N$ particles of $\{\theta_{0}^{i}\}_{i=1,...,N} \sim p(\theta_{0}) $. Note that $\theta_{t}$ denotes estimates for the hyper-parameters after observing $x_{1:t}$, and not that $\theta$ is time-varying.
   \STATE Set initial importance weights, $w_{0}^{i}=\frac{1}{N}$
   \FOR{$t=1$ {\bfseries to} $T$}
		\STATE Compute point estimates $m_{t}^{i}$ and $\mu_{t}^{i}$ \label{lst:line:apf1}
		\begin{align}
			m_{t}^{i} &= a \theta _{t-1}^{i} + (1-a) \bar{\theta}_{t-1} \label{eq:shrinkageKernel} \\
			\bar{ \theta}_{t-1} &= \frac{1}{N} \sum_{n=1}^{N} \theta_{t-1}^{n}\notag  \\
			\mu_{t}^{i}  &= \mathbb{E}(v_{t}| m_{t}^{i}, v_{1:t-1}^{i}, x_{1:t-1}) \label{eq:gpVolAPFProp}
		\end{align} 
		\STATE Compute point estimate importance weights: \label{lst:line:apf2}
		\begin{align} 
		 g_{t}^{i} \propto w_{t-1}^{i} p(x_{t} | \mu_{t}^{i}, m_{t}^{i} ) \label{eq:pointEstimateImportanceWeight}
		\end{align}
		\STATE Resample $N$ auxiliary indices $\{j\}$ according to weights $\{g_{t}^{i}\}$. \label{lst:line:apf3}
		\STATE Propagate these chains forward, i.e. set $\{ v_{1:t-1}^{i}\}_{i=1}^{N}$=$\{ v_{1:t-1}^{j}\}_{j \in J}$.
		\STATE Jitter the parameters $\theta_{t}^{j} \sim \mathcal{N}(m_{t}^{j}, (1-a^{2}) V_{t-1}) $, where $V_{t-1}$ is the empirical covariance of $\theta_{t-1}$. \label{lst:line:shrinkageKernel}
		\STATE Propose new states $v_{t}^{j} \sim  p(v_{t} | \theta_{t}^{j}, v_{1:t-1}^{j}, x_{1:t-1}^{j})$ 
		\STATE Compute the importance weights adjusting for the modified proposal: 
		\begin{align}
			w_{t}^{j} \propto \frac{ p(x_{t} | v_{t}^{j}, \theta_{t}^{j} )} { p(x_{t} | \mu_{t}^{j}, m_{t}^{j} )} \label{eq:realImportanceWeights}
		\end{align} 
   \ENDFOR
   \STATE {\bfseries Output:} posterior particles for chain of states $v_{1:T}^{j}$, parameters $\theta_{t}^{j}$ and particle weights $w_{t}^{j}$.
\end{algorithmic}
\end{algorithm}
PGAS is described in Algorithm~\ref{alg:pgas}.
There are three main parts to PGAS.
First, it adopts a Gibbs sampling approach and alternatively samples the parameters $\theta[m]$ given all the data $x_{1:T}$ and current samples of the hidden states $v_{1:T}[m-1]$, where $m$ is the iteration count, and then the states given the new parameters and the data.
Conditionally sampling $\theta[m]$ given $x_{1:T}$ and $v_{1:T}[m-1]$ can often be done with slice sampling \citep{NealSlice2003}.
Second, samples for $v_{1:T}[m]$ are drawn using a conditional auxiliary particle filter with ancestor sampling (CAPF-AS).
CAPF-AS consists of two parts: a) conditional auxiliary particle filter (CAPF) and b) ancestor sampling.
CAPF generates $N$ particles for each hidden state conditional on $\theta[m]$ and $v_{1:T}[m-1]$.
The conditional dependence is necessary, as each hidden state $v_{t}$ depends on the parameters and all the other hidden states.
In particular \citet{lindsten2012ancestor} verified that the CAPF corresponds to a collapsed Gibbs sampling \citep{van2008partially}.
Then ancestor sampling is used to sample smoothed trajectories from the particles generated by CAPF.
Third, the alternate sampling of $\theta[m]$ and $v_{1:T}[m]$ is repeated for $M$ iterations.
\begin{algorithm}[H]
   \caption{PGAS}
   \label{alg:pgas}
\belowdisplayskip=-5pt 
\begin{algorithmic}[1]
   \STATE {\bfseries Input:} data $x_{1:T}$, number of particles $N$, number of iterations $M$, initial hidden states $v_{1:T}[0]$ and initial parameters $\theta[0]$.
   \FOR{$m=1$ {\bfseries to} $M$}
		\STATE Draw $\theta[m] \sim p(\cdot| v_{1:T}[m-1], x_{1:T})$. This can often be done by slice sampling.
		\STATE Run a CAPF-AS with $N$ particles, targeting $p(v_{1:T} | \theta[m], x_{1:T})$ conditional on $v_{1:T}[m-1]$.
		\STATE Draw a chain $v_{1:T}^{\star} \propto \{w_{T}^{i}\}_{i=1,...,N}$ .
		\STATE Set $v_{1:T}[m]=v_{1:T}^{\star}$.
   \ENDFOR
   \STATE {\bfseries Output:} Chains of particles and parameters $\{v_{1:T}[m], \theta_{t}[m]\}_{m=1:M}$.
\end{algorithmic}
\end{algorithm}
Experiments comparing RAPCF against PGAS for GP-Vol are included in Section~\ref{sec:expInf}.

\section{Experiments} \label{sec:expGPVol}
We performed three sets of experiments.
First, we tested whether we can learn the states and transition dynamics for the GP-Vol model using RAPCF.
This was done by generating synthetic data and having RAPCF recover the hidden states and transition dynamics. This experiment and results are described in detail in Section~\ref{sec:expGPVolSynthetic}.
Second, we compared the performance of GP-Vol against standard econometric models GARCH, EGARCH and GJR-GARCH on twenty real financial time series in Section~\ref{sec:expGPVolReal}.
Model performance was measured in terms of predictive log-likelihoods.
GP-Vol was on average the most predictive model.
Finally, we compared the performance of the RAPCF algorithm against PGAS in terms of predictive log-likelihoods and execution times in Section~\ref{sec:expInf}.

\subsection{Synthetic data} \label{sec:expGPVolSynthetic}
Ten synthetic datasets of length $T=100$ were generated according to Equations~\eqref{eq:gpVol0} and \eqref{eq:gpVol1}.
The function $f$ was specified with a linear mean function and a squared exponential covariance function.
The linear mean function used was:
\begin{align}
	\mathbb{E}(v_t) = m(v_{t-1},x_{t-1}) &= av_{t-1} + b x_{t-1} \label{eq:synMeanFunc} \,.
\end{align}
This mean function encodes an asymmetric relationship between the positive and negative returns and the volatility.
The squared exponential kernel or covariance function is given by:
\begin{align}
	k(y,z) &= \sigma_{f}^{2}\exp(-\frac{1}{2l^{2}} |y-z|^{2}) \label{eq:squareExpKernel} \,.
\end{align}
where $l$ is the length scale parameter and $\sigma_{f}^{2}$ is the signal variance parameter.

\begin{figure}[htbp]
  \begin{minipage}[b]{.4\linewidth}
    \centering
    \includegraphics[scale=.25]{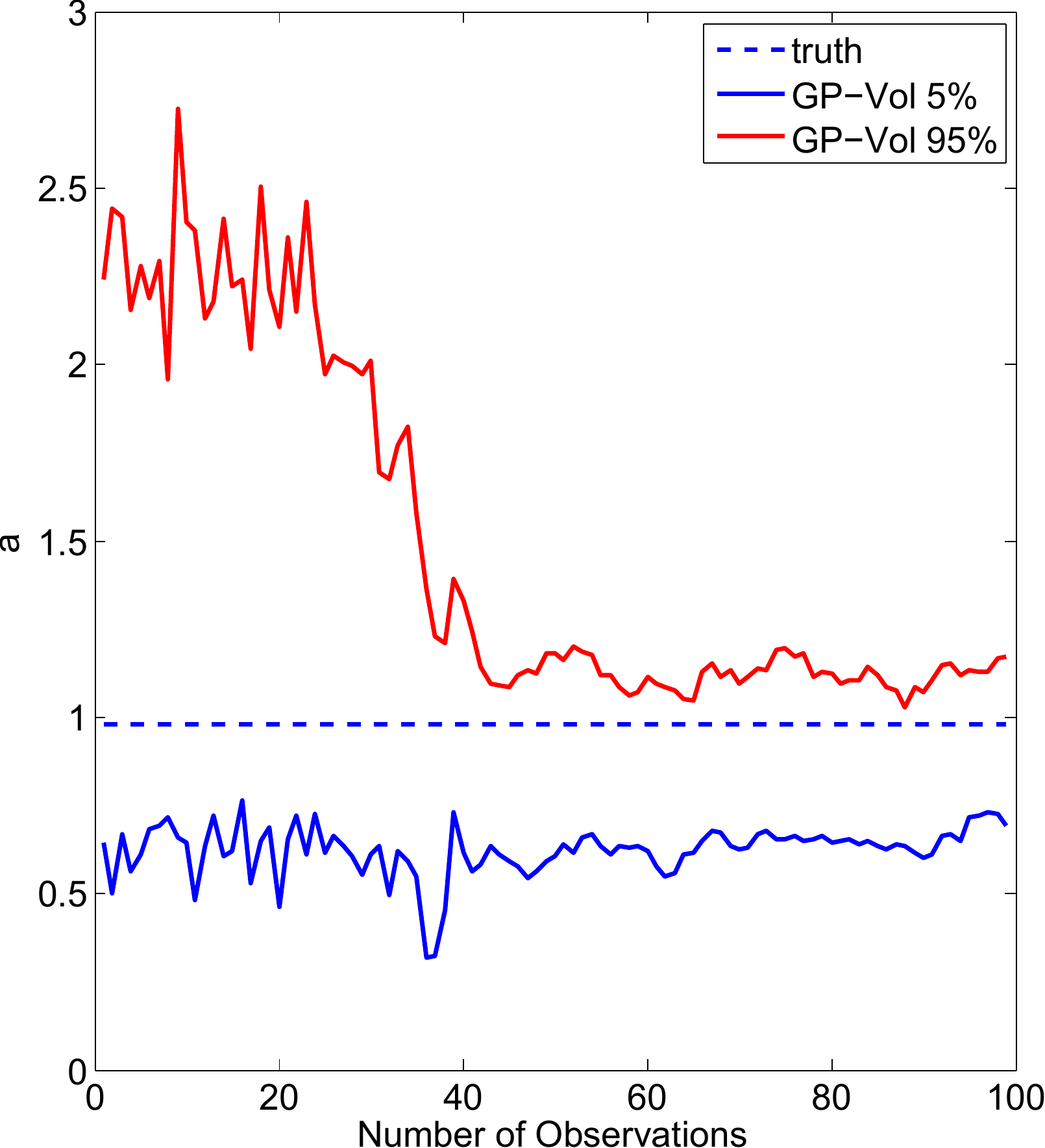} 
    \caption{90\% posterior interval for $a$.}
    \label{fig:synParamA}
  \end{minipage}
  \hspace{0.5cm}
  \begin{minipage}[b]{.4\linewidth}
    \centering
    \includegraphics[scale=.25]{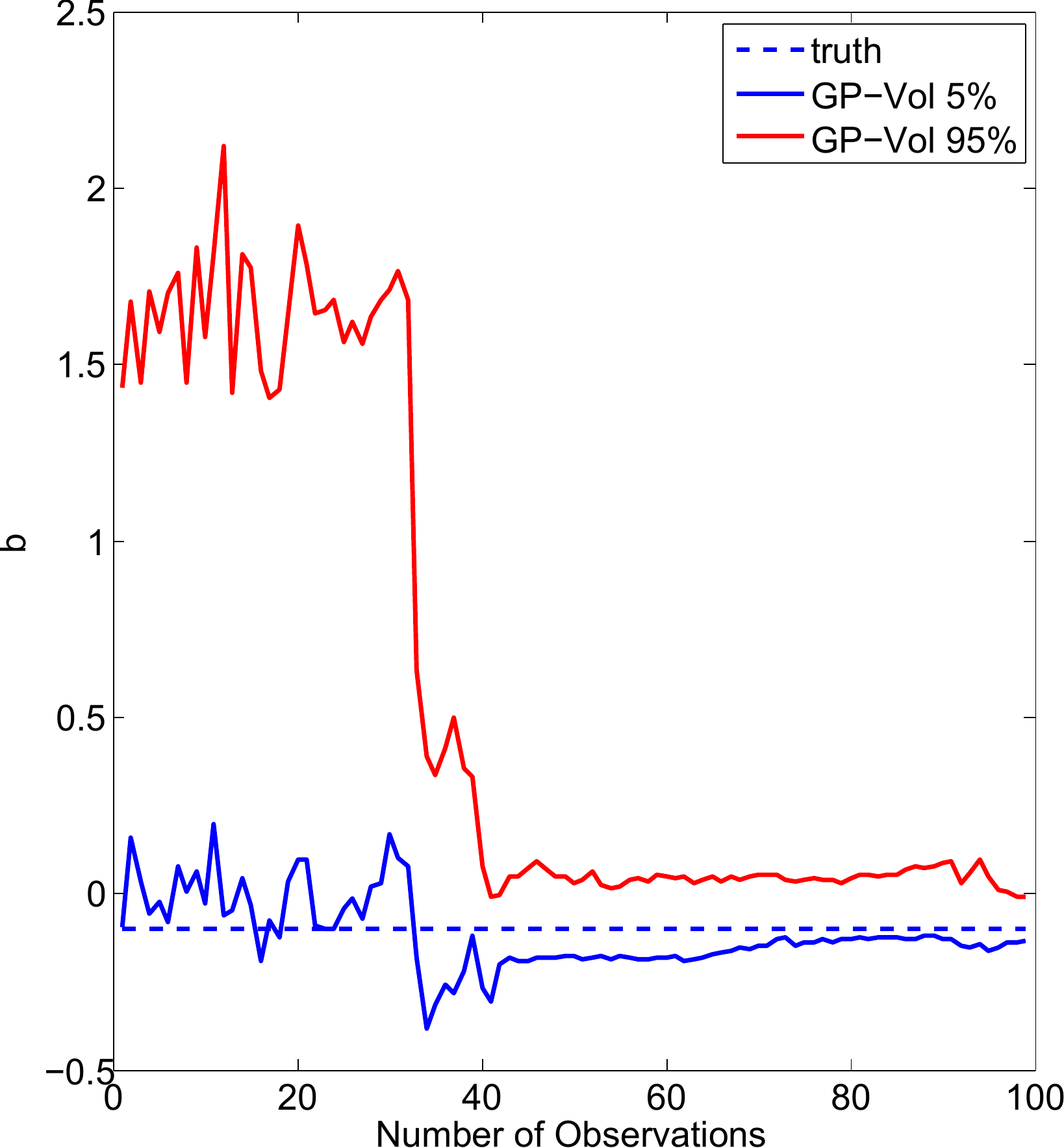} 
    \caption{90\% posterior interval for $b$.}
    \label{fig:synParamB}
  \end{minipage}
\end{figure}

RAPCF was used to learn the hidden states $v_{1:T}$ and the hyper-parameters $\theta=(a, b, \sigma_{n},\sigma_{f}, l)$ from $f$.
The algorithm was initiated with diffuse priors placed on $\theta$.
RAPCF was able to recover the hidden states and the hyper-parameters.
For the sake of brevity, we only include two typical plots of the $90\%$ posterior intervals for hyper-parameters $a$ and $b$ in Figures~\ref{fig:synParamA} and \ref{fig:synParamB} respectively.
The intervals are estimated from the filtered particles for $a$ and $b$ at each time step $t$.
In both plots, the posterior intervals eventually cover the parameters, shown as dotted blue lines, that were used to generate the synthetic dataset.

\subsection{Real data} \label{sec:expGPVolReal}

Experiments comparing GP-Vol, GARCH, EGARCH and GJR-GARCH were conducted on real financial datasets.
For these experiments, GARCH(1,1), EGARCH(1,1) and GJR-GARCH(1,1,1) were used, as they had the lowest number of parameters, thereby least susceptible to overfitting.
The financial datasets consisted of twenty time series of daily foreign exchange (FX) prices.
Each time series contained a total of $T=780$ observations from January 2008 to January 2011. 
The price data $p_{1:T}$ were pre-processed to eliminate spurious prices. 
In particular, we eliminated prices corresponding to times when markets were closed or not liquid.
Next the price data was converted into returns, $x_{t}=\log(p_{t}/p_{t-1})$.
Finally the returns were standardized to have zero mean and unit standard deviation.

The performance of each model is measured in terms of the predictive log-likelihood on the first return out of the training set.
During the experiments, each method receives an initial time series of length $100$.
The different models are trained on that data and then a one-step forward prediction is made.
The predictive log-likelihood is evaluated on the next observation out of the training set.
Then the training set is augmented with the new observation and the training and prediction steps are repeated.
The process is repeated sequentially until no further data is received.

GARCH, EGARCH and GJR-GARCH were implemented using numerical optimization routines provided by Kevin Sheppard
\footnote{\url{http:///www.kevinsheppard.com/wiki/UCSD\_GARCH/}}.
A relatively long initial time series of $100$ was needed to to train these models,
as using shorter initial data resulted in wild jumps in the maximum likelihood estimates of model parameters.
The large fluctuations in parameter estimates produced poor one-step forward predictions.

On the other hand, GP-Vol is less susceptible to overfitting as it approximates the posterior distribution using RAPCF instead of finding maximum likelihood point estimates.
For the experiments on real data, diffuse priors were placed on $\theta=(a,b,\sigma_{n},\sigma_{f},l)$, where $a$ and $b$ are the coefficients of a linear mean function, $\sigma_{n}$ is the process noise, and $\sigma_{f}$ ,  and $l$ the parameters of a squared exponential covariance function.
Finally, $N=200$ particles were used in the RAPCF.

Results showing the average predictive log-likelihood of GP-Vol, GARCH, EGARCH and GJR-GARCH are provided in Table~\ref{tbl:gpVolPredLiks}.
\newcommand{\ica}{\hspace{0.1cm}}
\renewcommand{\arraystretch}{0.85}
\begin{table}[h]
\centering
\caption{{Average predictive log-likelihood}}
\label{tbl:gpVolPredLiks}
\resizebox{0.45 \textwidth }{!}{
\begin{tabular}{@{\ica}l@{\ica}c@{\ica}c@{\ica}c@{\ica}c@{\ica}c@{\ica}}
Dataset &  GARCH   &  EGARCH   &  GJR   &  GP-Vol   \\ 
  \hline
AUDUSD & $-1.3036$ & $-1.5145$ & $-1.3053$ & $\mathbf{-1.2974}$ \\
BRLUSD & $-1.2031$ & $-1.2275$ & $-1.2016$ & $\mathbf{-1.1805}$ \\
CADUSD & $-1.4022$ & $-1.4095$ & $-1.4028$ & $\mathbf{-1.3862}$ \\
CHFUSD & $-1.3756$ & $-1.4044$ & $-1.4043$ & $\mathbf{-1.3594}$ \\
CZKUSD & $-1.4224$ & $-1.4733$ & $\mathbf{-1.4222}$ & $-1.4569$ \\
EURUSD & $-1.4185$ & $-2.1205$ & $-1.4266$ & $\mathbf{-1.4038}$ \\
GBPUSD & $\mathbf{-1.3827}$ & $-3.5118$ & $-1.3869$ & $-1.3856$ \\
IDRUSD & $-1.2230$ & $-1.2443$ & $-1.2094$ & $\mathbf{-1.0399}$ \\
JPYUSD & $-1.3505$ & $-2.7048$ & $-1.3556$ & $\mathbf{-1.3477}$ \\
KRWUSD & $-1.1891$ & $-1.1688$ & $-1.2097$ & $\mathbf{-1.1541}$ \\
MXNUSD & $-1.2206$ & $-3.4386$ & $-1.2783$ & $\mathbf{-1.1673}$ \\
MYRUSD & $-1.3940$ & $-1.4125$ & $-1.3951$ & $\mathbf{-1.3925}$ \\
NOKUSD & $-1.4169$ & $-1.5674$ & $-1.4190$ & $\mathbf{-1.4165}$ \\
NZDUSD & $\mathbf{-1.3699}$ & $-3.0368$ & $-1.3795$ & $-1.3896$ \\
PLNUSD & $-1.3952$ & $-1.3852$ & $\mathbf{-1.3829}$ & $-1.3932$ \\
SEKUSD & $-1.4036$ & $-3.7058$ & $\mathbf{-1.4022}$ & $-1.4073$ \\
SGDUSD & $\mathbf{-1.3820}$ & $-2.8442$ & $-1.3984$ & $-1.3936$ \\
TRYUSD & $\mathbf{-1.2247}$ & $-1.4617$ & $-1.2388$ & $-1.2367$ \\
TWDUSD & $-1.3841$ & $-1.3779$ & $-1.3885$ & $\mathbf{-1.2944}$ \\
ZARUSD & $-1.3184$ & $-1.3448$ & $\mathbf{-1.3018}$ & $-1.3041$ \\
\hline
\end{tabular}
}
\end{table}

\begin{figure}
	\centering \includegraphics[scale=.45]{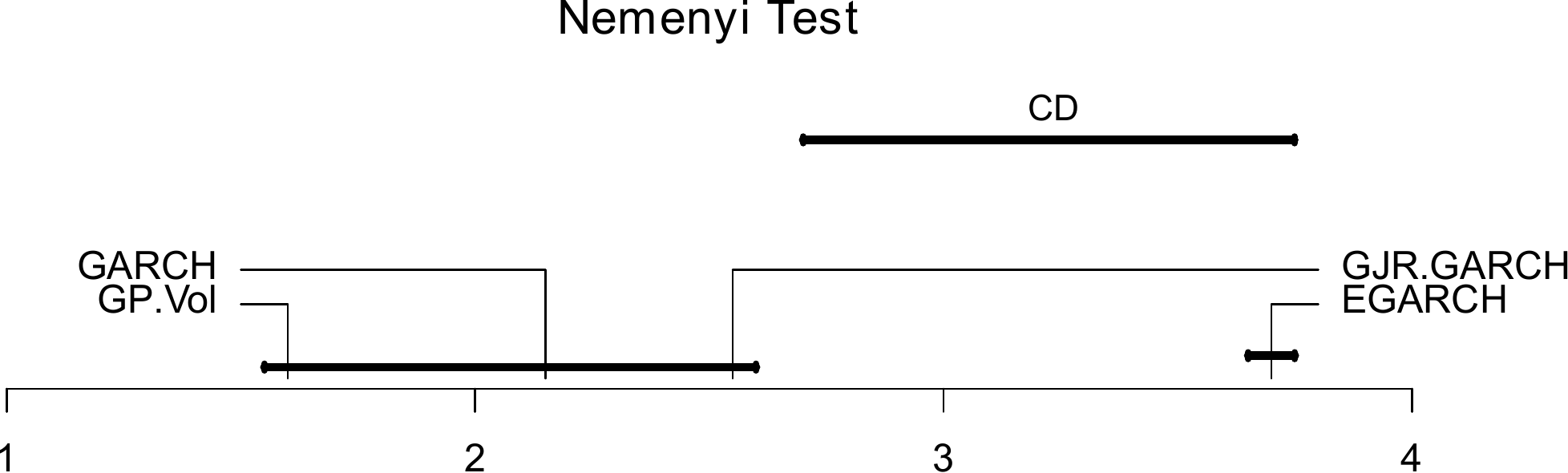} 
	\caption{All to all comparison between GP-Vol, GARCH, EGARCH and GJR-GARCH via a Nemenyi test.
	The horizontal axis indicates the average rank of each method on the 20 time series.
	If the differences in average ranks are larger than the critical distance (length of the segment labeled CD)
	then differences in performance are statistically significant at $\alpha = 0.05$.}
	\label{fig:modelRanks}
\end{figure}

The table shows that GP-Vol has the highest predictive log-likelihood in twelve of the twenty datasets.
We perform a statistical test to determine whether differences among GP-Vol, GARCH, EGARCH and GJR-GARCH are significant.
The methods are compared against each other using the multiple comparison approach described by \citet{demšar2006statistical}.
In this comparison framework, all the methods are ranked according to their performance on different tasks.
Statistical tests are then applied to determine whether the differences
among the average ranks of the methods are significant.
In our case, each of the $20$ datasets analyzed represents a different task.
Pairwise comparisons between all the methods with a Nemenyi test at a 95\% confidence level are summarized in Figure~\ref{fig:modelRanks}.
The methods whose average ranks across datasets differ more than a critical distance (segment labeled CD in the figure) show significant differences in performance at this confidence level.
The Nemenyi test shows that GP-Vol is the top-ranked model, but is not statistically superior to GARCH and GJR-GARCH at $\alpha=0.05$.
Note however that the Nemenyi test is very conservative.
In addition, the predictive performance of GARCH, EGARCH and GJR-GARCH are flattered by having a large initial time series of $100$, which mitigates overfitting from using maximum likelihood.
Even then EGARCH overfits and significantly underperforms the other models.

While GP-Vol did not dominate all the other models on all the tasks,
pairwise comparisons of GP-Vol to the other models via a Wilcoxon signed-rank test show significant outperformance at $\alpha=0.10$.
The p-values of these pairwise comparisons are given in Table~\ref{tbl:gpVolPredLiks}.
\begin{table}[h]
\centering
\caption{{
p-values for Wilcoxon signed-rank test of GP-Vol against the other three models.
}}
\label{tbl:gpVolPredLiks}
\resizebox{0.45 \textwidth }{!}{
\begin{tabular}{p{2cm}p{1.5cm}p{1.5cm}p{1.5cm}}
GP-Vol vs. &  GARCH   &  EGARCH   &  GJR    \\
  \hline
p-value & 0.079 & 0.0001 & 0.100 \\
\hline
\end{tabular}
}
\end{table}

\begin{figure}[htbp]
    \centering
    \includegraphics[scale=.5]{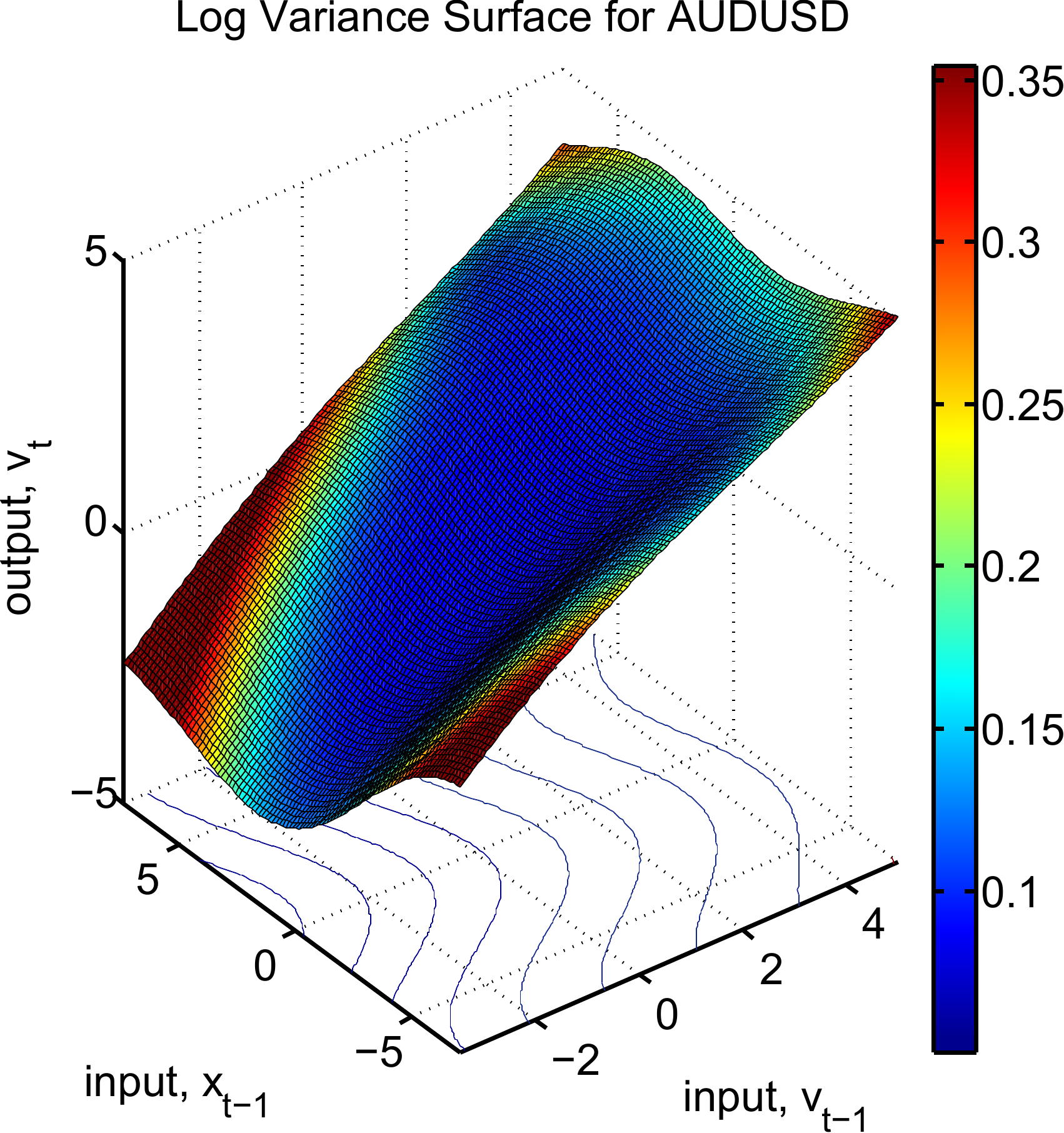} 
    \caption{Mean $f$ predictions for pairs of inputs $(v_{t-1},x_{t-1})$, colored according to prediction standard deviations.}
    \label{fig:gpVolSurf}
\end{figure}

The other advantage of GP-Vol over existing models is that it learns the functional relationship $f$ between the 
log variance $v_{t}$ and the previous log variance and return $(v_{t-1}, x_{t-1})$.
We plot a typical log variance surface, Figure~\ref{fig:gpVolSurf}.
Here the surface is generated by plotting the mean predicted outputs $v_{t}$ against a grid of inputs $(v_{t-1},x_{t-1})$, given the functional dynamics learned on the AUDUSD time series.
AUDUSD stands for the amount of US dollars that an Australian dollar can buy.
The grid of inputs was designed to contain a range of values experienced by AUDUSD from 2008 to 2011.
In this highly volatile period, large standard deviations $\sigma_{t-1}=\exp(\max(v_{t-1})/2) \approx \exp(2)=7.4$ were experienced.
Similarly, large swings in returns $x_{t-1} \geq \abs{5}$ occurred.
The surface is colored according to the standard deviations of the predictions.
Large standard deviations correspond to uncertain predictions, and are redder.
Figure~\ref{fig:gpVolSurf} shows four main patterns.
First, there is an asymmetric effect of positive and negative previous returns $x_{t-1}$.
This can be seen in both the shape of the log variance surface and the skew of the contour lines.
Large, positive $x_{t-1}$ augurs lower next step log variance $v_{t}$.
Second, the relationship between $v_{t-1}$ and $v_{t}$ is not linear,
because the contour lines are not parallel along the $v_{t-1}$ axis.
In addition, the relationship between $x_{t-1}$ and $v_{t-1}$ is nonlinear, but some sort of skewed quadratic function.
These two patterns confirm the asymmetric effect and the nonlinear transition function that EGARCH and GJR-GARCH attempt to model.
Third, there is a dip in predicted log variance for $v_{t-1}<-2$ and $-1<x_{t-1}<2.5$.
Intuitively this makes sense, as it corresponds to a calm market environment with low volatility.
However, as $x_{t-1}$ becomes more extreme the market becomes more turbulent, and $v_{t}$ increases.
Finally and non-intuitively, $v_{t}$ decreases as $x_{t-1}$ increases except in high volatility markets with $v_{t-1}>4$.
Digging into the data, we see that in those environments, large previous returns are often bounce-backs from large negative returns.
Therefore the asset is still experiencing a period of high volatility.

\begin{figure}[htbp]
  \begin{minipage}[b]{.4\linewidth}
    \centering
    \includegraphics[scale=.25]{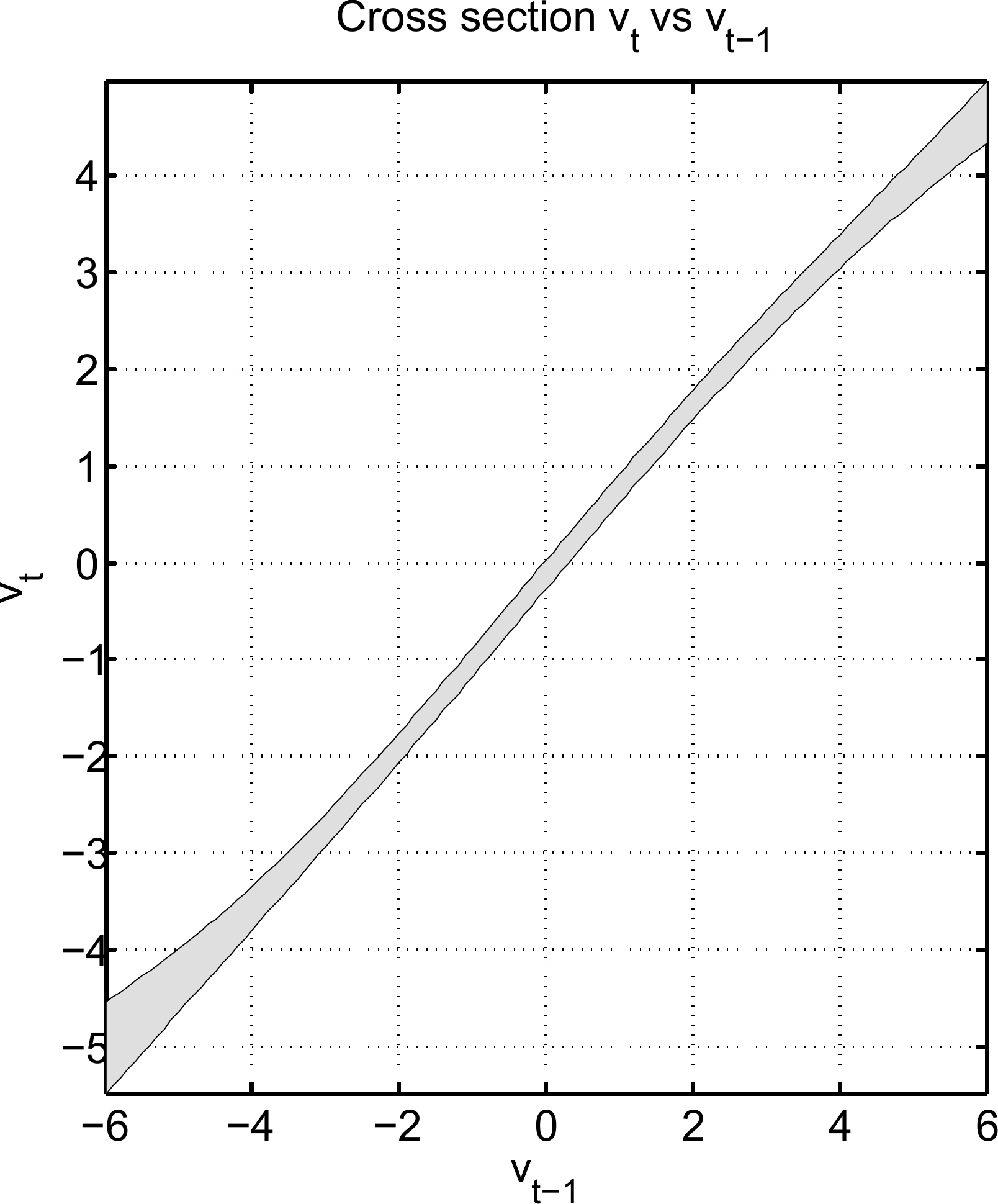} 
    \caption{Predicted $v_{t} \pm 2$ s.d. for inputs $(v_{t-1},0)$}
    \label{fig:gpVolSurfV2}
  \end{minipage}
  \hspace{0.5cm}
  \begin{minipage}[b]{.4\linewidth}
    \centering
    \includegraphics[scale=.25]{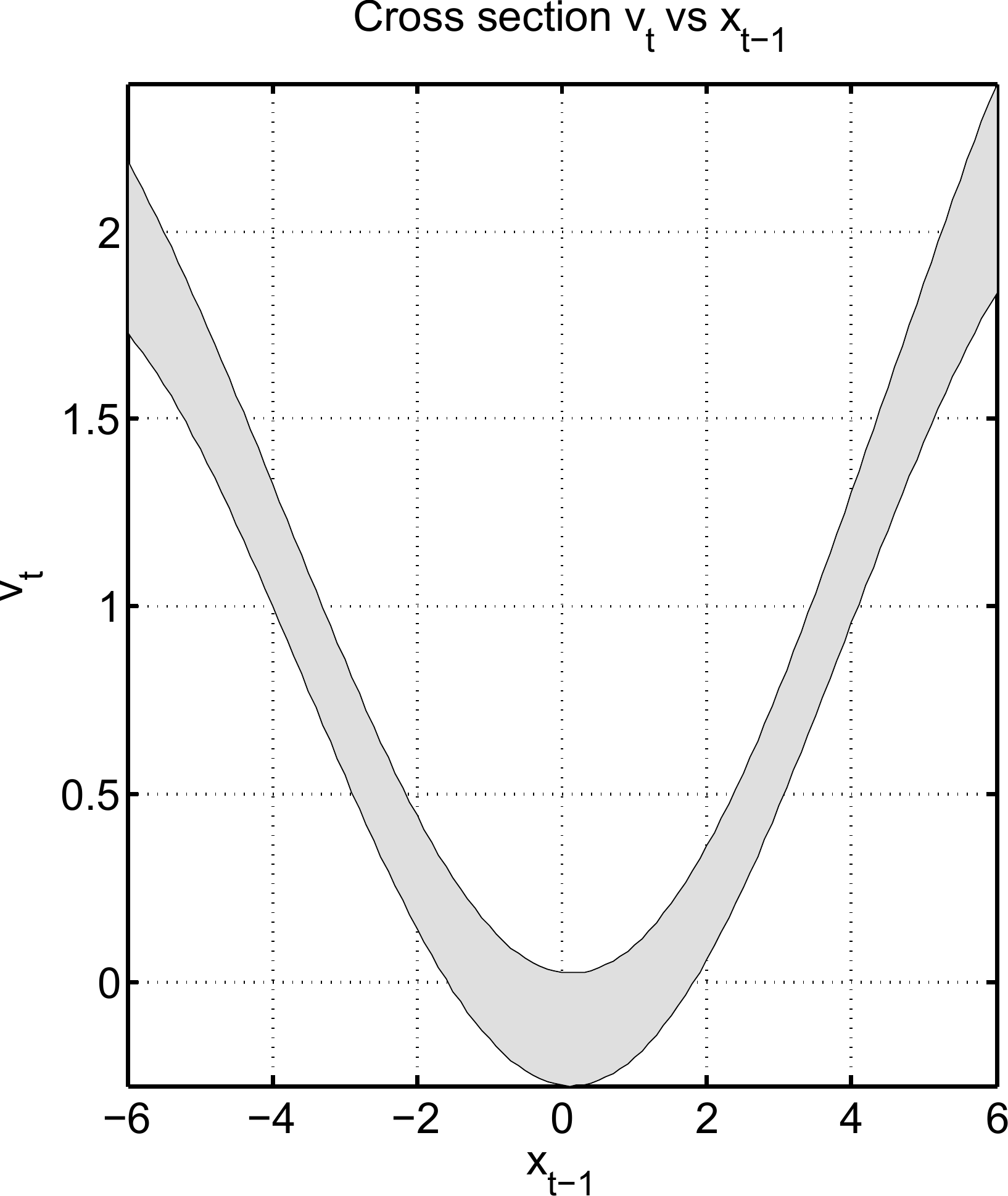} 
    \caption{Predicted $v_{t} \pm 2$ s.d. for inputs $(0,x_{t-1})$}
    \label{fig:gpVolSurfX2}
  \end{minipage}
\end{figure}

To further understand the transition function $f$, we study cross sections of the log variance surface.
First, $v_{t}$ is predicted for a grid of $v_{t-1}$ and zero $x_{t-1}$ in Figure~\ref{fig:gpVolSurfV2}.
Next, predicted $v_{t}$ for various $x_{t-1}$ and zero $v_{t-1}$ is shown in Figure~\ref{fig:gpVolSurfX2}.
The bands in the figures correspond to the mean prediction $\pm 2 $ standard deviations.
%

The cross sections confirm the nonlinearity of the transition function and the asymmetric effect of positive and negative returns on log variance.
Note that the transition function on $v_{t-1}$ looks linear, but is not as the band passes through $(-2,-2)$ and $(0,0)$, but not $(2,2)$ in Figure~\ref{fig:gpVolSurfV2}.

\subsection{RAPCF vs PGAS} \label{sec:expInf}
To understand the potential shortcomings of RAPCF discussed in Section~\ref{sec:GPinferenceMethods}, we compare it against PGAS on the twenty financial time series in terms of predictive log-likelihood and execution times.
The RAPCF setup is the same as in Section~\ref{sec:expGPVolReal}.
For PGAS, which is a batch method, the algorithm is run on initial training data $x_{1:L}$, with $L=100$,
and a one-step forward prediction is made.
The predictive log-likelihood is evaluated on the next observation out of the training set.
Then the training set is augmented with the new observation and the training and prediction steps are repeated.
The process is repeated sequentially until no further data is received.
For these experiments we used shorter time series with $T=120$, as PGAS was expensive computationally.
Note that we cannot simply learn the GP-SSM dynamics on a small set of training data and predict for a large test dataset as in \citet{frigola2013bayesian}.
In \citet{frigola2013bayesian}, the authors were able to predict forward as they were using synthetic data with known ``hidden'' states.

Different settings of RAPCF and PGAS were compared.
The setting for RAPCF was fixed to have $N=200$ particles since that was used to compare against GARCH, EGARCH and GJR-GARCH.
For PGAS, which has two parameters: a) $N$, the number of particles and b)  $M$, the number of iterations,
three combinations of settings were used.
The average predictive log-likelihood for RAPCF and PGAS are shown in Table~\ref{tbl:gpVolRAPFvsPGAS}.
\begin{table}[h]
\caption{{Average predictive log-likelihood}}
\label{tbl:gpVolRAPFvsPGAS}
\resizebox{0.45 \textwidth }{!}{
\begin{tabular}{@{\ica}l@{\ica}c@{\ica}c@{\ica}c@{\ica}c@{\ica}c@{\ica}}
Dataset &  RAPCF   &  PGAS.1   &  PGAS.2  &  PGAS.3   \\ 
 - &  N=200 &  N=10, M=100 &  N=25, M=100 &  N=10, M=200   \\
  \hline
AUDUSD & $-1.1205$ & $\mathbf{-1.0571}$ & $-1.0699$ & $-1.0936$ \\
BRLUSD & $-1.0102$ & $-1.0043$ & $-0.9959$ & $\mathbf{-0.9759}$ \\
CADUSD & $-1.4174$ & $-1.4778$ & $-1.4514$ & $\mathbf{-1.4077}$ \\
CHFUSD & $\mathbf{-1.8431}$ & $-1.8536$ & $-1.8453$ & $-1.8478$ \\
CZKUSD & $-1.2263$ & $-1.2357$ & $-1.2424$ & $\mathbf{-1.2093}$ \\
EURUSD & $-1.3837$ & $-1.4586$ & $\mathbf{-1.3717}$ & $-1.4064$ \\
GBPUSD & $-1.1863$ & $-1.2106$ & $-1.1790$ & $\mathbf{-1.1729}$ \\
IDRUSD & $-0.5446$ & $\mathbf{-0.5220}$ & $-0.5388$ & $-0.5463$ \\
JPYUSD & $-2.0766$ & $\mathbf{-1.9286}$ & $-2.1585$ & $-2.1658$ \\
KRWUSD & $\mathbf{-1.0566}$ & $-1.1212$ & $-1.2032$ & $-1.2066$ \\
MXNUSD & $-0.2417$ & $-0.2731$ & $\mathbf{-0.2271}$ & $-0.2538$ \\
MYRUSD & $\mathbf{-1.4615}$ & $-1.5464$ & $-1.4745$ & $-1.4724$ \\
NOKUSD & $-1.3095$ & $-1.3443$ & $\mathbf{-1.3048}$ & $-1.3169$ \\
NZDUSD & $-1.2254$ & $\mathbf{-1.2101}$ & $-1.2366$ & $-1.2373$ \\
PLNUSD & $-0.8972$ & $\mathbf{-0.8704}$ & $-0.8708$ & $\mathbf{-0.8704}$ \\
SEKUSD & $\mathbf{-1.0085}$ & $\mathbf{-1.0085}$ & $-1.0505$ & $-1.0360$ \\
SGDUSD & $\mathbf{-1.6229}$ & $-1.9141$ & $-1.7566$ & $-1.7837$ \\
TRYUSD & $\mathbf{-1.8336}$ & $-1.8509$ & $-1.8352$ & $-1.8553$ \\
TWDUSD & $\mathbf{-1.7093}$ & $-1.7178$ & $-1.8315$ & $-1.7257$ \\
ZARUSD & $\mathbf{-1.3236}$ & $-1.3326$ & $-1.3440$ & $-1.3286$ \\
\hline
\end{tabular}
}
\end{table}

\begin{figure}
	\centering \includegraphics[scale=.45]{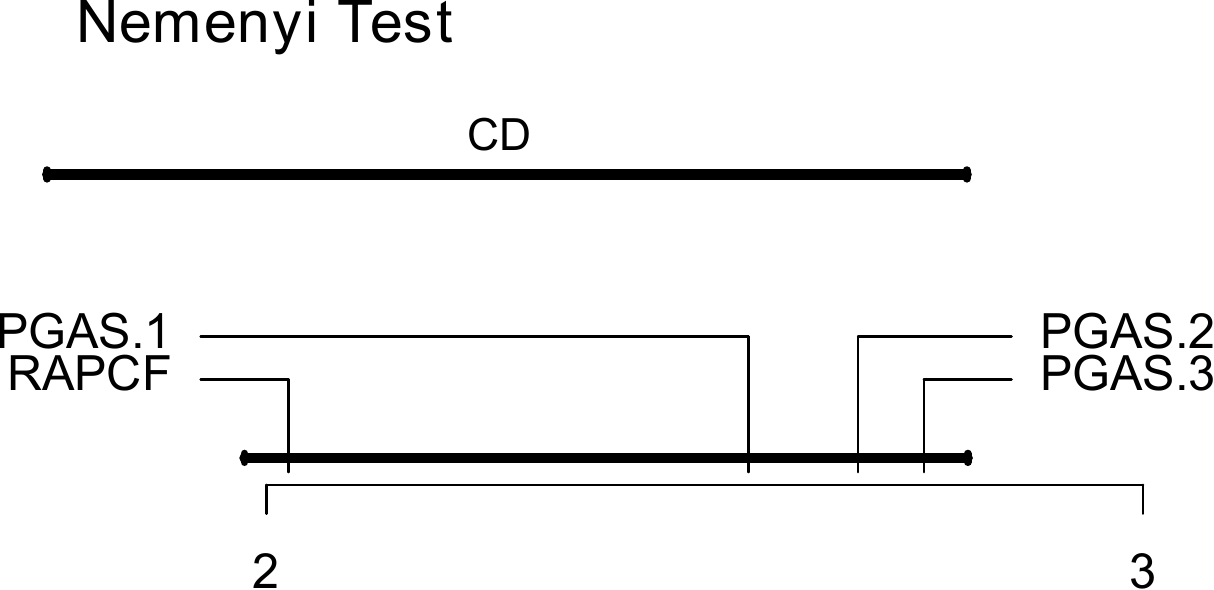} 
	\caption{All to all comparison between RAPCF and the three PGAS settings via a Nemenyi test.}
	\label{fig:infRanks}
\end{figure}

From the table there is no evidence that PGAS outperforms RAPCF on financial datasets,
since there is no clear predictive edge of any PGAS setting over RAPCF on the twenty time series.
A Nemenyi test at 90\% confidence level for the four inference methods is summarized in Figure~\ref{fig:infRanks}.
It shows no significant differences between the average predictive ranks of the inference methods.

While there is little difference in prediction accuracy between RAPCF and PGAS, PGAS is much more expensive computationally.
Average execution times for RAPCF and PGAS on the twenty financial datasets of length $T=120$ are shown in Table~\ref{tbl:gpVolRAPFvsPGASTimes}.
\begin{table}[h]
\caption{{Run time in minutes }}
\label{tbl:gpVolRAPFvsPGASTimes}
\resizebox{0.45 \textwidth }{!}{
\begin{tabular}{@{\ica}l@{\ica}c@{\ica}c@{\ica}c@{\ica}c@{\ica}c@{\ica}}
 \hline
Avg Time  &  RAPCF   &  PGAS   &  PGAS   &  PGAS   \\
 - &  N=200&  N=10, M=100 &  N=25, M=100&  N=10, M=200   \\
  \hline
Min & $6$ & $732$ & $1832$ & $1465$ \\
\hline
\end{tabular}
}
\end{table}

Of course PGAS can be calibrated to use fewer particles or iterations, but PGAS will still be more expensive than RAPCF.
A naive implementation of RAPCF will have $O(NT^{4})$, since at each time step $t$ there is a $O(T^{3})$ cost of inverting the covariance matrix.
On the other hand, the complexity of applying PGAS naively is $O(NMT^{5})$,
since for each batch of data $x_{1:t}$ there is a $O(NMT^{4})$ cost.
These costs can be reduced to be $O(NT^{3})$ and $O(NMT^{4})$ for RAPCF and PGAS respectively by doing rank one updates of the inverse of the covariance matrix at each time step.
The costs can be further reduced by a factor of $T^{2}$ with sparse GPs \citep{quinonero2005unifying}.

\section{Summary} \label{sec:summaryChap6}

We have introduced a novel Gaussian Process Volatility Model (GP-Vol) model for time-varying variances.
GP-Vol is an instance of a Gaussian Process State-Space model (GP-SSM).
It is highly flexible and can model nonlinear functional relationships and asymmetric effects of positive and negative returns on time-varying variances.
In addition, we have presented an online inference method based on particle filtering for GP-Vol.
This inference method is much quicker than the current batch Particle Gibbs method, and can be more generally applied to other GP-SSMs.
Results for GP-Vol on real financial data show significant predictive improvement over existing models such as GARCH, EGARCH and GJR-GARCH.
Finally, the nonlinear function representations learned using GP-Vol is highly intuitive with clear financial explanations.

There are two main directions for future work.
First, GP-Vol can be extended to learn the functional relationship between a financial instrument's volatility, its price and other market factors, such as interest rates.
The functional relationship thus learned will be useful in the pricing of volatility derivatives on the instrument.
Second, the speed of RAPCF makes it an attractive choice for live tracking of complex control problems.

{\normalfont \small
\bibliography{example_paper}

\begin{thebibliography}{27}
\providecommand{\natexlab}[1]{#1}
\providecommand{\url}[1]{\texttt{#1}}
\expandafter\ifx\csname urlstyle\endcsname\relax
  \providecommand{\doi}[1]{doi: #1}\else
  \providecommand{\doi}{doi: \begingroup \urlstyle{rm}\Url}\fi

\bibitem[Andrieu et~al.(2010)Andrieu, Doucet, and
  Holenstein]{andrieu2010particle}
Andrieu, Christophe, Doucet, Arnaud, and Holenstein, Roman.
\newblock {Particle Markov chain Monte Carlo methods}.
\newblock \emph{Journal of the Royal Statistical Society: Series B (Statistical
  Methodology)}, 72\penalty0 (3):\penalty0 269--342, 2010.

\bibitem[Baum \& Petrie(1966)Baum and Petrie]{baum1966statistical}
Baum, L.E. and Petrie, T.
\newblock Statistical inference for probabilistic functions of finite state
  {M}arkov chains.
\newblock \emph{The Annals of Mathematical Statistics}, 37\penalty0
  (6):\penalty0 1554--1563, 1966.

\bibitem[Bekaert \& Wu(2000)Bekaert and Wu]{bekaert2000asymmetric}
Bekaert, Geert and Wu, Guojun.
\newblock Asymmetric volatility and risk in equity markets.
\newblock \emph{Review of Financial Studies}, 13\penalty0 (1):\penalty0 1--42,
  2000.

\bibitem[Bollerslev(1986)]{bollerslev1986generalized}
Bollerslev, T.
\newblock Generalized autoregressive conditional heteroskedasticity.
\newblock \emph{Journal of econometrics}, 31\penalty0 (3):\penalty0 307--327,
  1986.

\bibitem[Campbell \& Hentschel(1992)Campbell and Hentschel]{campbell1992no}
Campbell, John~Y and Hentschel, Ludger.
\newblock No news is good news: An asymmetric model of changing volatility in
  stock returns.
\newblock \emph{Journal of financial Economics}, 31\penalty0 (3):\penalty0
  281--318, 1992.

\bibitem[Cont(2001)]{Cont2001}
Cont, R.
\newblock Empirical properties of asset returns: Stylized facts and statistical
  issues.
\newblock \emph{Quantitative Finance}, 1\penalty0 (2):\penalty0 223--236, 2001.

\bibitem[Deisenroth \& Mohamed(2012)Deisenroth and
  Mohamed]{deisenroth2012expectation}
Deisenroth, Marc and Mohamed, Shakir.
\newblock {Expectation Propagation in Gaussian Process Dynamical Systems}.
\newblock In \emph{Advances in Neural Information Processing Systems 25}, pp.\
  2618--2626, 2012.

\bibitem[Deisenroth et~al.(2009)Deisenroth, Huber, and
  Hanebeck]{deisenroth2009analytic}
Deisenroth, Marc~Peter, Huber, Marco~F, and Hanebeck, Uwe~D.
\newblock Analytic moment-based {G}aussian process filtering.
\newblock In \emph{Proceedings of the 26th annual international conference on
  machine learning}, pp.\  225--232. ACM, 2009.

\bibitem[Dem{\v{s}}ar(2006)]{demšar2006statistical}
Dem{\v{s}}ar, J.
\newblock Statistical comparisons of classifiers over multiple data sets.
\newblock \emph{Journal of Machine Learning Research}, 7:\penalty0 1--30, 2006.

\bibitem[Doucet et~al.(2001)Doucet, De~Freitas, and
  Gordon]{doucet2001sequential}
Doucet, A., De~Freitas, N., and Gordon, N.
\newblock \emph{{Sequential Monte Carlo methods in practice}}.
\newblock Springer Verlag, 2001.

\bibitem[Engle(1982)]{engle1982autoregressive}
Engle, R.F.
\newblock Autoregressive conditional heteroscedasticity with estimates of the
  variance of {U}nited {K}ingdom inflation.
\newblock \emph{Econometrica: Journal of the Econometric Society}, pp.\
  987--1007, 1982.

\bibitem[Frigola et~al.(2013)Frigola, Lindsten, Sch\"{o}n, and
  Rasmussen]{frigola2013bayesian}
Frigola, Roger, Lindsten, Fredrik, Sch\"{o}n, Thomas~B., and Rasmussen, Carl~E.
\newblock Bayesian inference and learning in {G}aussian process state-space
  models with particle {MCMC}.
\newblock In Bottou, L., Burges, C.J.C., Ghahramani, Z., Welling, M., and
  Weinberger, K.Q. (eds.), \emph{Advances in Neural Information Processing
  Systems 26}, pp.\  3156--3164. 2013.
\newblock URL
  \url{http://media.nips.cc/nipsbooks/nipspapers/paper_files/nips26/1449.pdf}.

\bibitem[Glosten et~al.(1993)Glosten, Jagannathan, and
  Runkle]{glosten1993relation}
Glosten, Lawrence~R, Jagannathan, Ravi, and Runkle, David~E.
\newblock On the relation between the expected value and the volatility of the
  nominal excess return on stocks.
\newblock \emph{The journal of finance}, 48\penalty0 (5):\penalty0 1779--1801,
  1993.

\bibitem[Hentschel(1995)]{hentschel1995all}
Hentschel, Ludger.
\newblock {All in the family nesting symmetric and asymmetric GARCH models}.
\newblock \emph{Journal of Financial Economics}, 39\penalty0 (1):\penalty0
  71--104, 1995.

\bibitem[Ko \& Fox(2009)Ko and Fox]{ko2009gp}
Ko, Jonathan and Fox, Dieter.
\newblock {GP-BayesFilters: Bayesian filtering using Gaussian process
  prediction and observation models}.
\newblock \emph{Autonomous Robots}, 27\penalty0 (1):\penalty0 75--90, 2009.

\bibitem[L{\'a}zaro-Gredilla \& Titsias(2011)L{\'a}zaro-Gredilla and
  Titsias]{Lazaro2011}
L{\'a}zaro-Gredilla, Miguel and Titsias, Michalis~K.
\newblock Variational heteroscedastic {G}aussian process regression.
\newblock In \emph{ICML}, pp.\  841--848, 2011.

\bibitem[Lindsten et~al.(2012)Lindsten, Jordan, and
  Sch{\"o}n]{lindsten2012ancestor}
Lindsten, Fredrik, Jordan, Michael, and Sch{\"o}n, Thomas.
\newblock {Ancestor Sampling for Particle Gibbs}.
\newblock In \emph{Advances in Neural Information Processing Systems 25}, pp.\
  2600--2608, 2012.

\bibitem[Liu \& West(1999)Liu and West]{liu1999combined}
Liu, J. and West, M.
\newblock \emph{Combined parameter and state estimation in simulation-based
  filtering}.
\newblock Institute of Statistics and Decision Sciences, Duke University, 1999.

\bibitem[Neal(2003)]{NealSlice2003}
Neal, Radford~M.
\newblock Slice sampling.
\newblock \emph{The Annals of Statistics}, 31\penalty0 (3):\penalty0 705--741,
  2003.
\newblock ISSN 00905364.
\newblock URL \url{http://www.jstor.org/stable/3448413}.

\bibitem[Nelson(1991)]{nelson1991conditional}
Nelson, D.B.
\newblock {Conditional heteroskedasticity in asset returns: A new approach}.
\newblock \emph{Econometrica}, 59\penalty0 (2):\penalty0 347--370, 1991.

\bibitem[Pitt \& Shephard(1999)Pitt and Shephard]{pitt1999filtering}
Pitt, M.K. and Shephard, N.
\newblock Filtering via simulation: Auxiliary particle filters.
\newblock \emph{Journal of the American Statistical Association}, pp.\
  590--599, 1999.

\bibitem[Qui{\~n}onero-Candela \& Rasmussen(2005)Qui{\~n}onero-Candela and
  Rasmussen]{quinonero2005unifying}
Qui{\~n}onero-Candela, Joaquin and Rasmussen, Carl~Edward.
\newblock A unifying view of sparse approximate gaussian process regression.
\newblock \emph{The Journal of Machine Learning Research}, 6:\penalty0
  1939--1959, 2005.

\bibitem[Rasmussen \& Williams(2006)Rasmussen and
  Williams]{rasmussen2006gaussian}
Rasmussen, C.E. and Williams, C.K.I.
\newblock \emph{{Gaussian processes for machine learning}}.
\newblock Springer, 2006.

\bibitem[Turner et~al.(2010)Turner, Deisenroth, and Rasmussen]{turner2010state}
Turner, Ryan~D, Deisenroth, Marc~P, and Rasmussen, Carl~E.
\newblock State-space inference and learning with {G}aussian processes.
\newblock In \emph{International Conference on Artificial Intelligence and
  Statistics}, pp.\  868--875, 2010.

\bibitem[Van~Dyk \& Park(2008)Van~Dyk and Park]{van2008partially}
Van~Dyk, David~A and Park, Taeyoung.
\newblock {Partially collapsed Gibbs samplers: Theory and methods}.
\newblock \emph{Journal of the American Statistical Association}, 103\penalty0
  (482):\penalty0 790--796, 2008.

\bibitem[Wilson \& Ghahramani(2010)Wilson and Ghahramani]{NIPS2010_0784}
Wilson, Andrew and Ghahramani, Zoubin.
\newblock Copula processes.
\newblock In Lafferty, J., Williams, C. K.~I., Shawe-Taylor, J., Zemel, R.S.,
  and Culotta, A. (eds.), \emph{Advances in Neural Information Processing
  Systems 23}, pp.\  2460--2468. 2010.

\bibitem[Wu et~al.(2013)Wu, Lobato, and Ghahramani]{wu2013dynamic}
Wu, Yue, Lobato, Jos{\'e} Miguel~Hern{\'a}ndez, and Ghahramani, Zoubin.
\newblock Dynamic covariance models for multivariate financial time series.
\newblock In \emph{Proceedings of the 30th International Conference on Machine
  learning}, volume~3, pp.\  558--566, 2013.

\end{thebibliography}
\bibliographystyle{icml2014}
}

\end{document}